\documentclass[twocolumn,showpacs,preprintnumbers,amsmath,amssymb]{revtex4}
\usepackage{graphicx}% Include figure files
\usepackage{dcolumn}% Align table columns on decimal point
\usepackage{bm}% bold math
\usepackage{epstopdf}
\usepackage{mathrsfs}
\usepackage{amssymb}
\usepackage{amsmath}

\def\bs{\boldsymbol}
\def\del{\partial}

\def\p{{\boldsymbol p}}
\def\pb{\bar {\boldsymbol p}}

\def\q{{\boldsymbol q}}

\def\k{{\boldsymbol k}}

\def\x{{\boldsymbol x}}

\def\bkappa{{\boldsymbol \kappa}}
\def\bbkappa{\bar{\boldsymbol \kappa}}

\def\qqb{{q\bar q}}

\newcommand{\beq}{\begin{eqnarray}}
\newcommand{\eeq}{\end{eqnarray}}
\newcommand{\be}{\begin{eqnarray*}}
\newcommand{\ee}{\end{eqnarray*}}

\begin{document}

%\preprint{APS/123-QED}

\title{Antiangular Ordering of Gluon Radiation in QCD Media}% Force line breaks with \\

\author{Yacine Mehtar-Tani}
% \email{}
\author{Carlos A. Salgado}%
\author{Konrad Tywoniuk}
\affiliation{%
Departamento de F\'isica de Part\'iculas,
Universidade de Santiago de Compostela, 
E-15782 Santiago de Compostela, 
Galicia, Spain
}%

\date{\today}% It is always \today, today,

\begin{abstract}
We investigate angular and energy distributions of medium-induced gluon emission off a quark-antiquark antenna in the framework of perturbative QCD as an attempt toward understanding, from first principles, jet evolution inside the quark-gluon plasma. In-medium color coherence between emitters, neglected in all previous calculations, leads to a novel mechanism of soft-gluon radiation. The structure of the corresponding spectrum, in contrast with known medium-induced radiation, i.e., off a single emitter, retains some properties of the vacuum case: in particular, it exhibits a soft divergence. However, as opposed to the vacuum, the collinear singularity is regulated by the pair opening angle, leading to a strict angular separation between vacuum and medium-induced radiation, denoted as {\it antiangular ordering}. We comment on the possible consequences of this new contribution for jet observables in heavy-ion collisions. 
%%%
%{\bf this sentence is the same as in the decoherence paper, shall we keep it?}
%%%
\end{abstract}

\maketitle

Jets in hadronic collisions have proven to be one of the most accurate tests of perturbative QCD. The showering of soft gluons off partons originating from hard processes gives rise to interference effects which facilitate the measurement of explicit non-Abelian features of the theory. In vacuum, such emissions exhibit soft and collinear logarithmic divergences which compensate the smallness of the strong coupling constant and have to be resummed \cite{bas83}. Because of interference effects, the striking feature of strong angular ordering of subsequent emissions arises \cite{Mueller:1981ex,Ermolaev:1981cm}. Naturally, such a restriction on the available phase space leads to a strong suppression of soft-gluon emissions, dubbed the humpbacked plateau, which has been confirmed by experiment.

Compared to the state-of-the-art jet measurements in proton-(anti)proton collisions, jet physics in heavy-ion collisions is still in its early stages. Indeed, the investigation of these new possibilities has only recently started at the Brookhaven National Laboratory Relativistic Heavy Ion Collider \cite{Putschke:2008wn}. The advent of the LHC experimental program, with a high capability of jet measurements even in high-multiplicity events, motivates a fresh look on possible novel features of medium effects on jets which can provide interesting tools to probe the nature of the quark-gluon plasma.

On the theory side, efforts to address the question of whether and how the jet evolution is altered by the presence of the quark-gluon plasma have been put forward in the past few years. Although different medium effects could lead to changes in the jet properties, the modification of the gluon radiation pattern is expected to be of main relevance. This modification is so far only known for the inclusive one-gluon radiation off a fast quark or gluon \cite{bdmps,zakharov,gyu00,Wiedemann:2000za}. 

Clearly, color coherence effects among the different partons in the cascade are not addressed in this setup since only a single emitter is considered, nor is the presence of an ordering variable for subsequent emissions. Naively, one would expect a weakening of these effects in the medium due to momentum exchanges and color randomization. We will see, however, that this is not the case and that color interference effects 
might lead to strong modifications of the jet structure. 

In the vacuum, a quark-antiquark ($q\bar q$) antenna provides a simple laboratory for the intrajet coherent cascade, encompassing, in particular, the key feature of angular ordering \cite{Dokshitzer:1991wu}. In this work, we focus on the medium-induced part of the radiation spectrum off a $\qqb$ pair in a color-singlet state. Our results show that in the soft limit the interaction with the medium implies a color rotation of the pair as a whole, thereby inducing emissions at large angles analogous to the color-octet emission pattern in vacuum. However, in contrast to the latter, the collinear singularities are cut off by the pair opening angle due to medium-induced destructive interferences, thus leading to a geometrical separation between vacuum and medium-induced radiation. As we show below, these general features persist to large gluon energies. This is what we denote {\it antiangular ordering} of medium-induced radiation.

In short, in addition to accounting for coherence effects among scattering centers, as in Refs.~\cite{bdmps,zakharov,gyu00,Wiedemann:2000za}, in this work we extend these previous approaches by including coherence between emitters.

We shall proceed within the framework of the classical Yang-Mills (CYM) equations, which holds for soft-gluon radiation  \cite{meh07}. Here the amplitude of the emission of a gluon with momentum $k\equiv(\omega,\vec k)$, $\omega$ being its energy and $\vec k$ its 3-momentum vector, is related to the  classical gauge field by the reduction formula 
\beq
\label{eq:redform}
{\cal M}_\lambda ^{a}({\vec k})=\lim_{k^2\to 0 } -k^2A^{a}_\mu(k)  \epsilon^\mu_\lambda({\vec k}) \,,
\eeq
where $\epsilon^\mu_\lambda({\vec k})$ is the gluon polarization vector. The gauge field $A^\mu\equiv A^{\mu,a}t^a$, where $t^a$ is the generator of SU(3) in the fundamental representation, is the solution of the CYM equations $[D_\mu,F^{\mu\nu}] = J^\nu$, with $D_\mu\equiv \del_\mu-ig A_\mu$ and $F_{\mu\nu}\equiv \del_\mu A_\nu-\del_\nu A_\mu-ig[A_\mu,A_\nu]$. The covariantly conserved current, i.e., $[D_\mu,J^\mu]=0$,  describes the projectiles. Furthermore, we shall set our calculation in the light-cone gauge: $A^+\equiv (A^0+A^3)/\sqrt{2}=0$. With this choice, only the transverse polarizations contribute to the cross section with the help of the completeness relation 
$\sum_{\lambda} \epsilon^i_\lambda(\epsilon_\lambda^j)^\ast=\delta^{ij}$, where $i(j)=1,2$.  

Let us first briefly review the vacuum emission pattern of gluon emission off a $q \bar q$ pair with momenta $p\equiv(E, \vec p)$ and $\bar p\equiv( \bar E, \vec{\bar p})$, respectively. The classical eikonalized current that describes the pair created at time $t_0=0$ reads $J^{\mu}_{(0)}=J^\mu_{q}+J^\mu_{\bar q}$, where $ J^{\mu,a}_{q} = g\frac{p^\mu}{E}~\delta^{(3)}({\vec x}-\frac{\vec p}{E}t)~\Theta(t) ~Q_{q}^a$, and analogously for the antiquark. Here, $Q_q=-Q_{\bar q}=Q$ is the quark and antiquark color charge, respectively, with $Q^2=C_F\equiv(N_c^2-1)/2N_c$. Because of the gauge choice it is suitable to use the light-cone variables, e.g., $k\equiv[k^+=(\omega+k^3)/\sqrt{2}$, $k^-=(\omega-k^3)/\sqrt{2},\k]$, $\k\equiv(k^1,k^2)$, and similarly for any vector in what follows. At leading order in $g$ the linearized  CYM equations, in the light-cone gauge, read in momentum representation
\beq
 -k^2 A_{(0)}^{i,a} (k) = -ig\ \left(\frac{\kappa^i}{x\,(p \cdot k)}  - \frac{\bar \kappa^i}{\bar x\, (\bar p\cdot k)} \right) Q^a,
\eeq
where we have introduced the transverse vectors $\kappa^i = k^i - x \,p^i$ and $\bar \kappa^i = k^i - \bar x\, \bar p^i$ ($i=1,2$) and momentum fractions $x = k^+/p^+$ and $\bar x = k^+/\bar p^+$. The soft-gluon emission amplitude is connected to the gauge field through the reduction formula in Eq.~(\ref{eq:redform}), and, by summing over the gluon polarization vectors, it can be easily checked that the cross section reads
\beq
\label{eq:spectrum-vac1}
(2\pi)^2\,\omega\frac{dN^\text{vac}}{d^3k}=\frac{\alpha_s C_F}{\omega^2} \frac{2\, n_q\cdot n_{\bar q}}{(n_q\cdot n)(n_{\bar q}\cdot n)} \,,
\eeq
where $n_q^\mu \equiv p^\mu/E$, $n^\mu_{\bar q}\equiv \bar p^\mu/\bar E$, and $n^\mu\equiv k^\mu/\omega$.  
The cross section in Eq.~(\ref{eq:spectrum-vac1}) exhibits an apparent double collinear singularity. The two poles can be split into two separate terms which comprise the quark and the antiquark collinear divergences, respectively.

Averaging the collinear singular part of Eq.~(\ref{eq:spectrum-vac1}) along the direction of, e.g., the quark over the azimuthal angle leads to gluon emission confined to a cone defined by the opening half-angle of the $\qqb$ pair, $\theta_{q\bar q}$. This owes to the fact that large-angle radiation is suppressed since it does not resolve the internal structure of the pair. Thus, the corresponding gluon emission probability off the quark in vacuum reads  
\beq
\label{eq:spectrum-vac2}
dN^{\text{vac}}_q=\frac{\alpha_sC_F}{\pi} \frac{d\omega}{\omega}\frac{\sin\theta \ d \theta}{1-\cos\theta} \Theta(\cos\theta-\cos\theta_{q\bar q}),
\eeq
which exhibits a double logarithmic singularity, namely, a soft divergence, when $\omega\to 0$, and a collinear divergence, when $\theta\to 0$, where $\theta$ is the angle between the quark and the emitted gluon.

We now return to the medium modification of the gluon spectrum given by Eq.~(\ref{eq:spectrum-vac2}). Hereafter, we assume that the $q\bar q$ pair moves in the $+z$ direction and interacts with a medium moving in the opposite direction at nearly the speed of light. At the end of the calculation we will boost back to the medium rest frame. Therefore, this approximation is valid as long as the pair opening angle $\theta_{q\bar q}\ll1$ and at asymptotic energies. Also, we restrict our calculation, for simplicity, to first order in opacity, i.e., two-gluon exchange with the medium at the level of the cross section. To do so, the pair field is treated as a perturbation around the strong medium field $A_\text{med}$. In the asymptotic limit, the medium gauge field can be described by $A^-_\text{med}(x^+,\x)$ which is a solution of the Poisson equation $-{\bs \del}^2 A^-_\text{med}(x^+,\x)=\rho_\text{med}(x^+, \x)$, where the medium source density $\rho_\text{med}$ is treated as a Gaussian white noise, while $A_\text{med}^i=A^+_\text{med}=0$ \cite{meh07}. In Fourier space it reads
\beq
A^-_\text{med}(q) = 2\pi \, \delta(q^+)\int_{0}^{\infty} \!\!\! dx^+ ~{\cal A}_\text{med}(x^+,\q)~e^{i q^-x^+}.
\eeq
The medium average is defined as
\begin{multline}
\langle {\cal A}^a_\text{med}(x^+,\q) {\cal A}^{\ast b}_\text{med}(x'^+,\q')\rangle\equiv \\
 \qquad \delta^{ab}\,n_0 \,m_D^2\,\delta(x^+-x'^+)\ (2\pi)^2 \,\delta^{(2)}(\q-\q'){\cal V}^2(\q),
\end{multline}
where ${\cal V}(\q)=1/(\q^2+m_D^2)$ is the Coulomb potential, $m_D$ is the Debye mass, and $n_0$ is the one-dimensional density of scattering centers. At first order in the medium field, the continuity relation reads $\del_\mu J_{(1)}^\mu = ig [A^-_\text{med},J_{(0)}^+]$, which is solved by 
\beq
J^\mu_{(1)} = ig\frac{p^\mu}{p\cdot \del }~[A^-_\text{med},J^+_q]+ig\frac{ \bar p^\mu}{ \bar p\cdot \del }~[A^-_\text{med},J^+_{\bar q}] \;.
\eeq
Then, the transverse part of the CYM equations reads
\beq
\label{eq:field1}
\square A_{(1)}^i =2ig\left[A_\text{med}^-,\del^+ A^i_{(0)} \right] -\frac{\del^i}{\del^+}J^+_{(1)}+J^i_{(1)} \;,
\eeq
which in momentum space becomes
\begin{multline}
\label{eq:field1}
\!\!\!\!\!-k^2A_{(1)}^i(k) = 2g\int \frac{d^4 q}{(2\pi)^4} (q-k)^+ \left[A_\text{med}^-(q),  A^i_{(0)}(k-q) \right] \\ -\frac{k^i}{k^+}J^+_{(1)}(k)+J^i_{(1)}(k) \;.
\end{multline}
After performing the $q^+$ and $q^-$ integrals we obtain the amplitude for gluon radiation off the quark, via Eq.~(\ref{eq:redform}):
\begin{multline}
\label{eq:amplitude}
{\cal M}^{a}_{q,(1)} =ig^2 \int  \frac{d^2\q}{(2\pi)^2}\int_{0}^{L^+} \!\!\!\!\! dx^+\, \left[T\cdot{\cal A}_\text{med}(x^+,\q)\right]^{ab}Q_q^b \\
\hspace{-2em}\times\left[ \frac{{\bs \nu}\cdot {\bs \epsilon}}{x\,(p\cdot v)} \left(1-e^{i\frac{p\cdot v}{p^+}x^+}\right)+\frac{{\bs \kappa}\cdot {\bs \epsilon}}{x\,(p\cdot k)}e^{i\frac{p\cdot v}{p^+}x^+}\right] \,,
\end{multline}
modulo a phase that cancels in the cross section, where $v \equiv \left[k^+,(\k-\q)^2\big/2k^+, \k-\q\right]$ and $\nu^i =v^i-x\,p^i$, and $T$ is the SU(3) generator in the adjoint representation. 
In Eq.~(\ref{eq:amplitude}), $L^+=\sqrt{2}L$, where $L$ is the medium size. 
The amplitude for gluon radiation off the antiquark  is deduced from $\mathcal{M}_q$ by the substitution $p \to \bar p$.
The first term  in Eq.~(\ref{eq:amplitude}) corresponds to the interaction of the emitted gluon with the medium, denoted ${\cal M }^{g}_q$, while the second term corresponds to gluon bremsstrahlung where only the quark interacts, denoted ${\cal M}_q^{\text{brem}}$. The contact terms, being the interference between the gluon emission amplitude in vacuum and the one accompanied by two-gluon scattering with the medium, are essential for unitarity and simply lead to a redefinition of the potential such that $ {\cal V}^2(\q)\to{\cal V}^2(\q)-\delta^{(2)}(\q)\int d^2\q' {\cal V}^2(\q')$, which guarantees that the spectrum is finite in the $\q \to0$ limit.

Squaring the amplitude and summing over the polarization vector, $|\mathcal{M}|^2=|\mathcal{M}_q|^2 + |\mathcal{M}_{\bar q}|^2 + 2\,\text{Re}\,\mathcal{M}_q \mathcal{M}_{\bar q}^\ast$, we recover the Gyulassy-Levai-Vitev (GLV) spectrum \cite{gyu00,Wiedemann:2000za} for the quark and the antiquark, respectively (first two terms on the right-hand side) or, equivalently, the first order in opacity of the Baier-Dokshitzer-Mueller-Peign\'e-Schiff-Wiedemann-Zakharov spectrum \cite{bdmps,zakharov,Wiedemann:2000za}. The sum of the two we denote by $\mathcal{I}_{\text{GLV}}$. Additionally, we also get novel contributions stemming from the interference. The latter can be further divided into two contributions, namely $\mathcal{I}_{\text{brems}} = 2\,\text{Re}\,{\cal M}^{\text{brem}}_q {\cal M}^{*\text{brem}}_{\bar q}$, which is the only term exhibiting a soft divergence, and the remaining ones, involving at least one gluon interaction with the medium, denoted by $\mathcal{I}_{\text{interf}}$.

\begin{figure}
\includegraphics[width=0.39\textwidth]{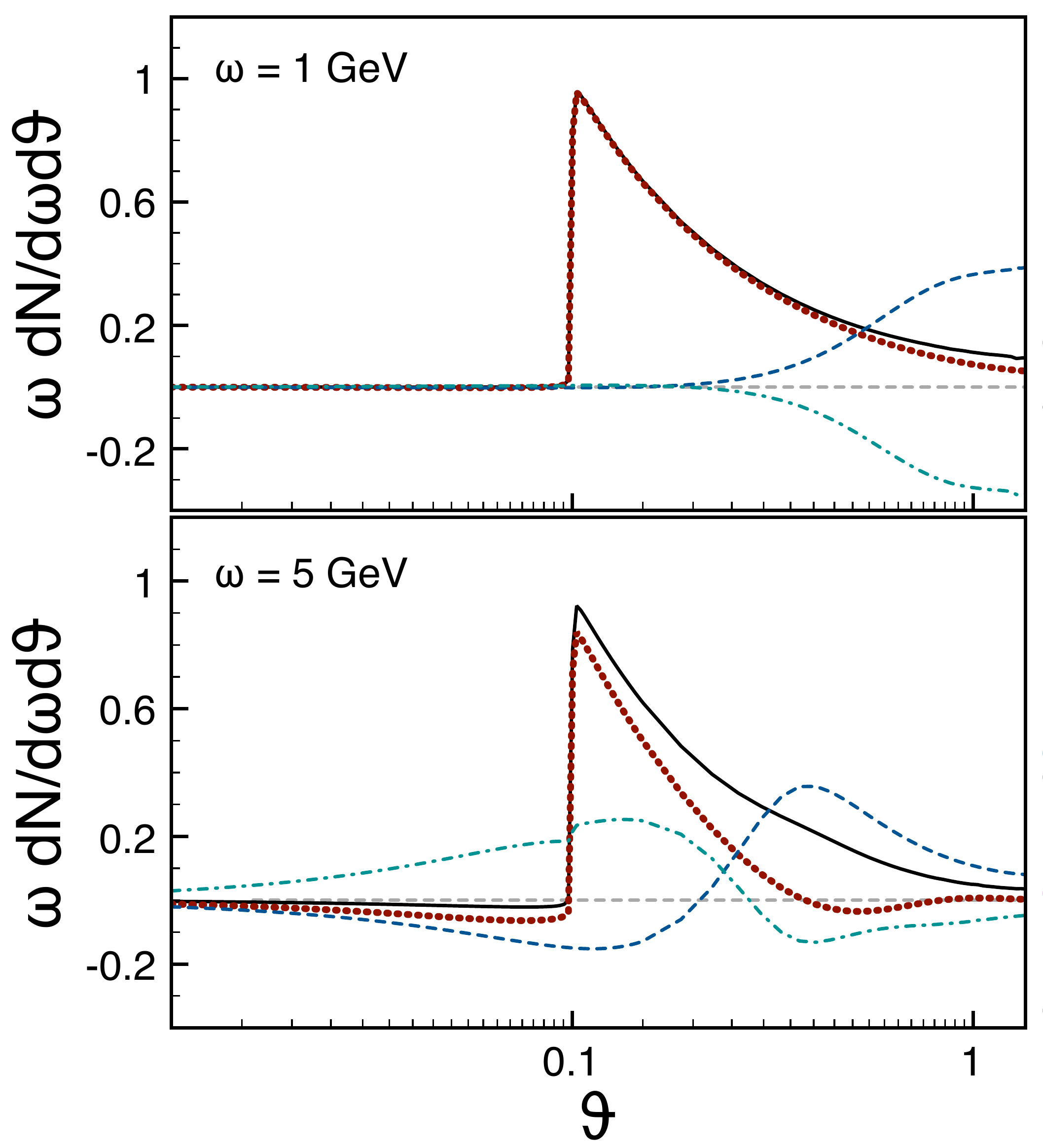}
\caption{The angular distribution of the medium-induced gluon spectrum for $\omega$=1 and 5 GeV for a $q\bar q$ pair with opening angle $\theta_{q\bar q}= 0.1$; see the text for details. The dotted (red) line corresponds to the dominant contribution in the soft limit, $\mathcal{I}_{\text{brems}}$ [see Eq. (\ref{eq:SpectrumMedSoft})], while the short-dashed (blue) curve is the sum of GLV contributions from the quark and the antiquark, $\mathcal{I}_{\text{GLV}}$, and the dash-dotted (green) curve depicts the remaining terms, $\mathcal{I}_{\text{interf}}$. The solid line corresponds to the total spectrum.}
\label{fig1}
\end{figure}

The three contributions are plotted in Fig. \ref{fig1}, where we have evaluated the angular distribution of the full spectrum of a $q \bar q$ pair with opening angle $\theta_{q\bar q} = 0.1$ traversing a medium with thickness $L = 4$ fm ($m_D=0.5$ GeV, $\alpha_s = 1/3$, and $n_0L^+=1$) numerically for two gluon energies. 
Here we have assumed the quark momentum to be along the $z$ axis for simplicity, i.e., $|\p|=0$.
We note that, in both cases, the three terms add up to zero at small angles, leaving the cone delimited by the pair angle free of radiation. The distribution jumps from zero inside the cone to a maximum value at $\theta=\theta_{q\bar q}$; it then drops as $ 1/\theta$ for $\theta>\theta_{q\bar q}$. This vacuumlike pattern persists at the higher energy (see Fig.~\ref{fig1} bottom) caused by an intricate cancellation between the different contributions and differs notably from the single-particle GLV spectrum.

In fact, these general features can be understood in the soft limit, i.e., $\omega\to 0$, where the dominating contribution to the spectrum is simply given by $\mathcal{I}_{\text{brems}}$, namely
\beq
\label{eq:SpectrumMedSoft}
\omega\frac{dN^{\text{med}}}{d^3k}&=& \frac{8\pi C_AC_F\, \alpha_s^2 \,n_0\, m_D^2}{(2\pi)^2} \frac{\bkappa \cdot \bbkappa}{x\,\bar x\, (p\cdot k)( \bar p\cdot k)} \\
&& \hspace{-5em}\times\int_0^{L^+}\!\!\!\! d x^+ \cos \Omega^0\,x^+ \int \frac{d^2\q }{(2\pi)^2} {\cal V}^2(\q) \left(1-\cos \Delta\Omega\, x^+ \right) , \nonumber
\eeq
where we have written the contact term explicitly and where $\Omega^0= p\cdot k/p^+ - \bar p \cdot k/\bar p^+$, $\Delta\Omega= \p\cdot \q/p^+ - \pb \cdot \q/\bar p^+$. The soft divergence in Eq.~(\ref{eq:SpectrumMedSoft}) is manifest. Note that in the soft limit, where $\Omega^0 \to 0$, the integrals in Eq.~(\ref{eq:SpectrumMedSoft}) are straightforward, yielding $ L^{+} r_\perp^2 [\ln \left(1/r_\perp m_D\right)+\text{const.}]/24\pi$, where  
$r_\perp=\theta_{q\bar q}L$.

Let us finally turn to the angular structure of Eq.~(\ref{eq:SpectrumMedSoft}). In the small angle limit $\theta_{q\bar q} \ll 1$ and $\theta \ll 1$, we get
\beq
{\cal J} = \frac{p^+\bar p^+ \, (\bkappa \cdot \bbkappa)}{(p\cdot k)(\bar p\cdot k)} \simeq -\frac{n_q\cdot n_{\bar q} - n_q\cdot n - n_{\bar q}\cdot  n}{(n_q\cdot n)(n_{\bar q}\cdot n)} \;,
\eeq
thus recovering a structure very similar to the vacuum one. 
As in the vacuum, we divide this term symmetrically between quark and antiquark. Then, averaging the quark contribution over the respective azimuthal angle, we obtain
\beq
\label{eq:AngularMed}
\frac{1}{2}\int \frac{d\varphi}{2\pi}\, {\cal J} = \frac{\Theta(\cos\theta_{q\bar q}-\cos\theta)}{1-\cos\theta} \;.
\eeq
The medium-induced soft-gluon radiation off the quark is suppressed inside the cone of opening angle $\theta_{q\bar q}$, as opposed to the standard angular structure obtained in vacuum; see Eq.~(\ref{eq:spectrum-vac2}). Furthermore, due to this feature, the collinear pole in Eq.~(\ref{eq:AngularMed}) is automatically cut off. 

Thus, when $\omega \rightarrow 0$ the medium-induced gluon emission off the quark can be written as
\beq
\label{eq:nqmed}
dN^{\text{med}}_q=\frac{\alpha_sC_F }{\pi}\Delta_{\text{med}}\frac{d\omega}{\omega}\frac{\sin\theta \ d \theta}{1-\cos\theta} \Theta(\cos\theta_{q\bar q}-\cos\theta),
\eeq
where $\Delta_{\text{med}} = \alpha_sC_A  n_0 m_D^2 L^{+} r_\perp^2 [\ln \left(1 \big/ r_\perp m_D\right)+\text{const.}] \linebreak/6$ is the forward dipole scattering amplitude in the adjoint representation.

The full gluon spectrum in the presence of a medium is thus given by $dN_q^{\text{tot}} = dN_q^{\text{vac}} + dN_q^{\text{med}}$. Equation~(\ref{eq:nqmed}) is the main result of this Letter, demonstrating, in particular, that there is a {\it strict geometrical separation} between vacuum and medium-induced radiation [cf. Eq.~(\ref{eq:spectrum-vac2})] and the appearance of a soft divergence in the latter.
\begin{figure}
\includegraphics[width=0.38\textwidth]{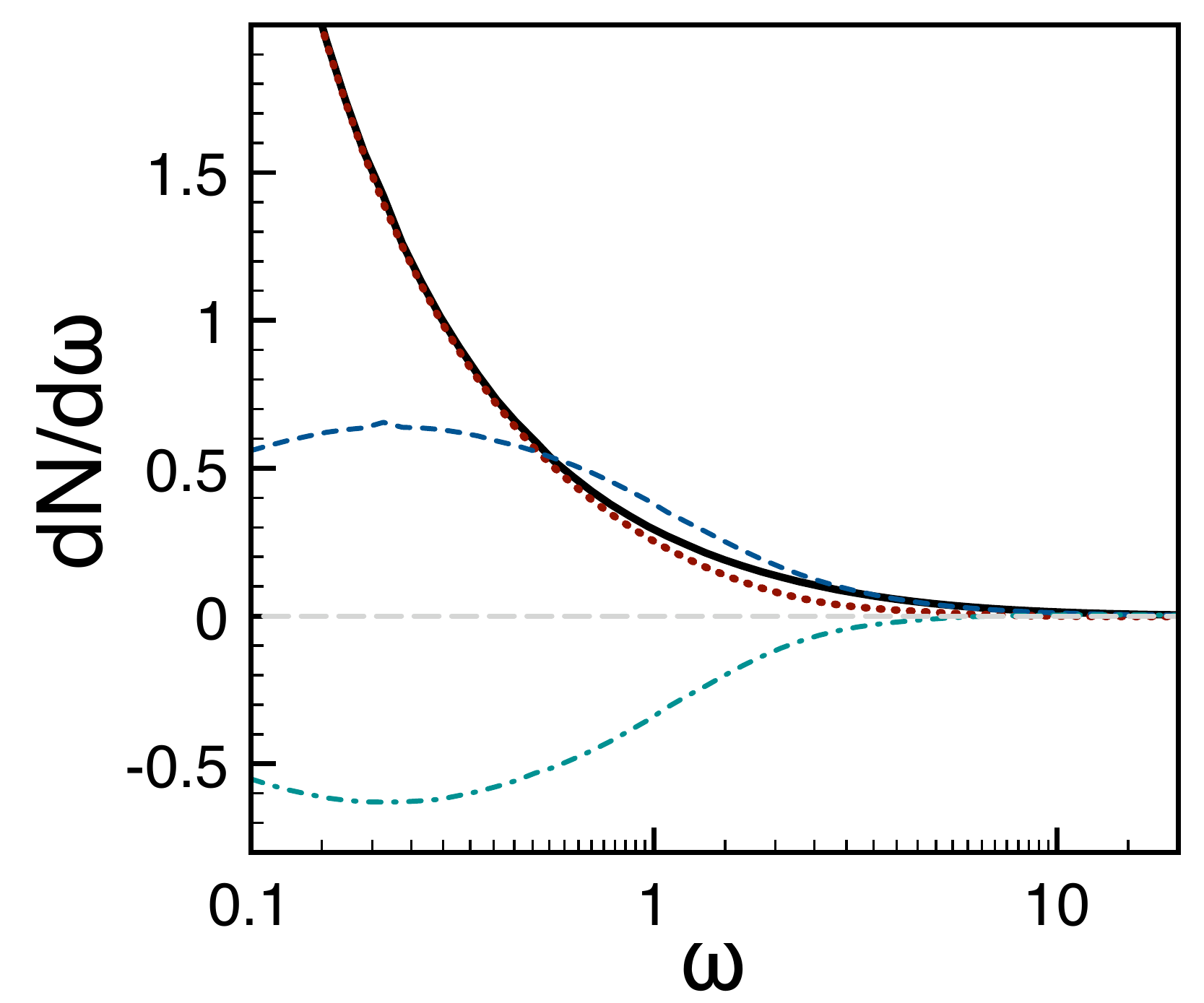}
\caption{The gluon spectrum integrated over the angle. Notations are the same as in Fig.~\ref{fig1}.}
\label{fig2}
\end{figure}

A natural cutoff appears in the bremsstrahlung interference spectrum due to the argument of the first cosine in Eq.~(\ref{eq:SpectrumMedSoft}): $\Omega^0 L^+\approx  \omega\theta_{q\bar q}^2 L \left(1/2-\theta/\theta_{q\bar q}\cos\varphi\right)$, which leads to an exponential drop of the spectrum at large $\omega$ driven by the new scale $1/\theta_{q\bar q}^2 L$. As a consequence, large opening angles $\theta_{q\bar q}$ as well as large medium sizes $L$ reduce the phase space for interference effects. 

For $\mathcal{I}_\text{GLV}$, gluons with formation times $t_f \sim 1/(\theta^2\omega) >L$  are suppressed due to the Landau-Pomeranchuk-Migdal effect \cite{bdmps,zakharov}, as shown in Fig.~\ref{fig1}. Consequently, the GLV spectrum is infrared and collinear safe \cite{sal03}. On the other hand, owing to the soft divergence the bremsstrahlung gluons, produced mostly at the angle $\sim \theta_\qqb$, have large formation times $t_f \gtrsim L$. These features are depicted in Fig.~\ref{fig2}, where we plot the gluon spectrum integrated over the angle (up to $\pi/2$), $dN\big/ d\omega$. To make contact with previous in-medium calculations, let us just remark that our contribution leads to a modest growth of about $30$\% (see Fig. \ref{fig2}) as compared to the standard GLV estimates. 

Let us now comment on the general structure of the medium-induced spectrum described above; cf. Eq.~(\ref{eq:nqmed}). Coherence among emitters occurs at large angles \cite{Dokshitzer:1991wu}. In the vacuum, this leads to a strong suppression due to destructive interferences with the incoherent spectra off the quark and antiquark, respectively. In the soft limit, the medium-induced spectrum off a single emitter, corresponding to the incoherent part $\mathcal{I}_\text{GLV}$, is suppressed due to formation time arguments, as mentioned before. Therefore, what remains is the coherent, large-angle emission. 

The simple vacuumlike form of the radiation spectrum in Eq.~(\ref{eq:nqmed}), persisting to large gluon energies due to an intricate interplay between various parts of the spectrum, is quite unexpected in light of the fact that the medium-induced spectrum off a single emitter is infrared safe and has an involved angular structure. Furthermore, in the soft limit it is easy to generalize these results to medium radiation off an antenna in a colored state. These circumstances point to a more general underlying mechanism of coherent gluon emission in medium and remain to be studied in detail.

In summary, we have calculated the gluon radiation spectrum off a quark-antiquark antenna immersed in a QCD medium, thus obtaining two crucial features: (i) a soft divergence present in the gluon bremsstrahlung term together with (ii) an antiangular ordering, which arises from medium-induced coherent radiation off multiple emitters. We note that the modification of the in-medium jet described above is of a different nature than the well-known broadening of the intrajet distribution \cite{Salgado:2003rv}. We expect this mechanism for soft-gluon radiation to be detectable for exclusive jet observables in heavy-ion collisions, in particular, as a nontrivial distortion of the humpbacked plateau in the soft sector. 

\begin{acknowledgments}
The authors thank N. Armesto, M. Braun, J. Casalderrey, Yu. Dokshitzer, A. Kovner and U. Wiedemann for stimulating discussions. This work is supported by Ministerio de Ciencia e Innovaci\'on of Spain; by Xunta de Galicia (Conseller\'{\i}a de Educaci\'on and Conseller\'\i a de Innovaci\'on e Industria -- Programa Incite); by the Spanish Consolider-Ingenio 2010 Programme CPAN; and by the European Commission. 
\end{acknowledgments}

\end{document}